\documentclass[twocolumn,showpacs,amsmath,amssymb,prl,aps,floats]{revtex4}
\usepackage{bm}
\usepackage{graphicx}
\usepackage{times}
\newcommand{\EQ}{\begin{equation}}
\newcommand{\EN}{\end{equation}}
\newcommand{\EQA}{\begin{eqnarray}}
\newcommand{\ENA}{\end{eqnarray}}

\newcommand{\Eq}[1]{Eq.~(\ref{#1})}

\newcommand{\xx}{\mbox{\boldmath $x$} {}}
\newcommand{\nn}{\mbox{\boldmath $n$} {}}
\newcommand{\kk}{\mbox{\boldmath $k$} {}}
\newcommand{\sss}{\mbox{\boldmath $s$} {}}

\begin{document}

\title{Cosmic Microwave Background bispectrum from primordial magnetic fields on large angular
scales}
\author{T. R. Seshadri}
\affiliation{Department of Physics and Astrophysics, University of Delhi,
Delhi 110007, India}
\email{trs@physics.du.ac.in}
\author{Kandaswamy Subramanian}
\affiliation{IUCAA, Post Bag 4, Ganeshkhind, Pune 411 007, India.}
\email{kandu@iucaa.ernet.in}

\date{\today}
\begin{abstract}
Primordial magnetic fields lead to non-Gaussian signals in the
Cosmic Microwave Background (CMB) even at the lowest order, as 
magnetic stresses, and the temperature anisotropy they induce,
depend quadratically on the magnetic field.
In contrast, CMB non-Gaussianity due to inflationary scalar perturbations
arise only as a higher order effect. We propose here
a novel probe of stochastic primordial magnetic fields
that exploits the characteristic CMB non-Gaussianity that they induce.
In particular, we 
compute the CMB bispectrum ($b_{l_{_1}l_{_2}l_{_3}}$) induced
by stochastic primordial fields on large angular scales. We find a typical value of
$l_1(l_1+1)l_3(l_3+1) b_{l_{_1}l_{_2}l_{_3}} \sim 10^{-22}$,
for magnetic fields of strength $B_0 \sim 3$ nano Gauss and with a nearly
scale invariant magnetic spectrum.
Current observational limits on the bispectrum
allow us to set upper limits on 
$B_0 \sim 35$ nano Gauss, 
which can be improved by including
other magnetically induced contributions to the bispectrum. 

\end{abstract}

\maketitle

Magnetic fields are ubiquitous in the universe but
their origin and evolution is still not fully understood. 
A popular paradigm is that the observed magnetic fields result
from the dynamo amplification of small seed fields. 
Astrophysical dynamos have been extensively studied \cite{BS05}
but are not without their difficulties. 
Another interesting alternative is that the observed large-scale
magnetic fields have their origin in the early Universe, 
arising perhaps during inflation or some other
phase transition \cite{TW88}.
If a primordial magnetic field with a present-day strength 
of a few nano Gauss is generated in the early universe, it 
would strongly affect the temperature and polarization anisotropies 
of the Cosmic Microwave Background Radiation.
Such anisotropies arise both due to metric perturbations
induced by magnetic stresses as well as fluid motions due to the
Lorentz force \cite{Bcmb,trsks01,SSB03}. 
We study here a new probe of primordial magnetic fields, namely, the possible
non-Gaussian signals they induce in the CMB.

Previous work on non-Gaussian signals from a cosmological
magnetic field has focused on a homogeneous magnetic
field \cite{hom}. Such a field can result in correlations
breaking spatial isotropy and also possibly result in 
north-south anisotropy on the CMB sky.
We focus here on stochastic primordial fields which
are statistically isotropic and homogeneous, and consider the
non-Gaussian signals they induce.

Non-Gaussianity of the temperature anisotropy of
the CMB has recently aroused considerable interest
in the context of inflationary models.
In such models,
linearized quantum fluctuations of the inflaton on
sub-Hubble
scales lead eventually to classical long wavelength
curvature perturbations.
The Gaussian statistics of the initial quantum fluctuations
lead to Gaussian statistics for the curvature
perturbations and the induced CMB anisotropies.  Non-Gaussianity can come in these models only
through higher order effects. The magnetic stresses on the other hand, depend
quadratically on the
field. Hence, even for the field with Gaussian probability distribution,
the magnetic stresses have necessarily a non-Gaussian character
even at the lowest order. This will naturally lead to non-Gaussianity
in the CMB anisotropies induced by primordial fields. Thus bounds on CMB
non-Gaussianity can strongly constrain primordial
magnetic fields, or lead to their detection.

The CMB bispectrum or the 3-point function of
the CMB anisotropies, if non-zero, provides an important
and  useful characterization of CMB non-Gaussianity.
It contains more information, for example, than the
single point probability distribution function (PDF).
We compute here the simplest contribution 
to the CMB bispectrum, namely that due to the magnetically induced 
Sachs-Wolfe type effects.

The magnetic field is assumed to be 
a Gaussian random field.
On galactic scales and above, any velocity induced by Lorentz forces 
is generally so small
that it does not lead to appreciable distortion of 
the initial field \cite{jedam}.
So, the magnetic field simply redshifts away as ${\bf B}(%
{\bf x},t)={\bf b}_{0}({\bf x})/a^{2}$, 
where, ${\bf b}_{0}$ is the
magnetic field at the present epoch (i.e. at $z=0$ or $a=1$)

The magnetically induced Sachs-Wolfe type 
contribution to the temperature anisotropies
given in \cite{giovannini,finelli}, which arises
on large-angular scales, can be expressed as,
\EQ
\frac{\Delta T}{T}(\nn) =  
{\cal R} ~ \Omega_B(\xx_0 -\nn D^*).
\EN
Here $\Omega_B({\bf x}) = {\bf B}^2({\bf x},t)/(8\pi \rho_\gamma(t)) 
={\bf b}_0^2({\bf x})/(8\pi \rho_0)$, 
where $\rho_\gamma(t)$ and $\rho_0$ are respectively the CMB energy densities at times $t$ and at the present epoch.
(Note that the $\Delta T/T$ given above is obtained for
on large-angular scales, just as the usual 
Sachs-Wolfe effect.) 
Ref. \cite{giovannini} gives an analytic estimate ${\cal R} = R_\gamma/20$
as the Sachs-Wolfe contribution, where
$R_\gamma \sim 0.6$ is the fractional contribution of radiation energy density 
towards the total energy density of the relativistic component.
Further, the unit vector ${\bf n}$ gives
the direction of observation and $D^*$ is the (angular diameter) 
distance to the surface of last scatter. 
We have assumed an instantaneous recombination which is a 
good approximation for large angular scales.
Note that there could also be additional integrated Sachs-Wolfe (ISW)
contributions to ${\cal R}$ \cite{giovannini}.

The temperature fluctuations can be expressed in terms of
the spherical harmonics to give 
$\Delta T(\nn)/T = \sum_{l m} a_{lm} Y_{lm}(\nn)$,
where
\EQ
a_{lm}= 4\pi \frac{1}{i^l}\int \frac{d^3 k}{(2\pi)^3} ~
{\cal R} ~ \hat\Omega_B(\kk) ~ j_l(kD^*) Y^*_{lm}(\hat \kk). \label{alm}
\EN
Here $\hat\Omega_B(\kk)$ is the Fourier transform of $\Omega_B({\bf x})$. 
Since $\Omega_B({\bf x})$ is quadratic in ${\bf b}_0({\bf x})$, we have
$\hat\Omega_B(\kk) =\frac{1}{(2\pi)^{3}}\int d^3s ~b_i(\kk+\sss)b^*_i(\sss)/(8\pi \rho_0)$,
where now $b_i(\kk)$ is the Fourier transform of ${\bf b}_0({\bf x})$.

A measure of non-Gaussianity in the CMB temperature anisotropy is its 3-point correlation function and is called the bispectrum, $B^{m_{_1}m_{_2}m_{_3}}_{l_{_1}l_{_2}l_{_3}}$. In terms of the ${a_{lm}}$'s it is given by,
\EQ
B^{m_{_1}m_{_2}m_{_3}}_{l_{_1}l_{_2}l_{_3}}=<a_{{l_1}{m_1}}a_{{l_2}{m_2}}a_{{l_3}{m_3}}>.
\EN
From Eq.~\ref{alm} we can express $B^{m_{_1}m_{_2}m_{_3}}_{l_{_1}l_{_2}l_{_3}}$ as
\EQ
B^{m_{_1}m_{_2}m_{_3}}_{l_{_1}l_{_2}l_{_3}}
= {\cal R}^3\int 
\left[\prod_{i=1}^3 (-i)^{l_i}
\frac{d^3k_i}{2\pi^2} j_{_{l_i}}(k_{_i}D^*)Y^*_{l_im_i}(\hat{\kk}_{_i})\right]
\zeta_{_{123}}
\label{bispec}
\EN
with $\zeta_{_{123}}$ defined as, 
\EQ
\zeta_{_{123}}=<\hat\Omega_B(\kk_{_1})\hat\Omega_B(\kk_{_2})
\hat\Omega_B(\kk_{_2})>.
\EN

The magnetic field itself is assumed to be non-helical and to have a Gaussian probability
distribution.
Hence it is completely specified by its energy power spectrum 
say $M(k)$. 
This spectrum is defined by the relation
$<b_{i}({\bf k})b^*_{j}({\bf q}%
)>=(2\pi)^3 \delta({{\bf k}-{\bf q}})P_{ij}({\bf k})M(k)$, where 
$P_{ij}({\bf k}) = (\delta_{ij} - k_ik_j/k^2)$ is the projection
operator ensuring ${\bf \nabla}\cdot{\bf b_0} =0$.
This gives $<{\bf b}_{0}^{2}>=2\int (dk/k)\Delta _{b}^{2}(k)$, where $%
\Delta _{b}^{2}(k)=k^{3}M(k)/(2\pi ^{2})$ is the power per logarithmic
interval in $k$ space residing in {the stochastic magnetic field}.

As in \cite{SSB03}, we assume a power-law magnetic spectra, 
$M(k)=Ak^{n}$ that has a cutoff at $k=k_{c}$,
where $k_{c}$ is the Alfv\'{e}n-wave damping length-scale
\cite{jedam}. We fix $A$ by demanding
that the {variance of the} magnetic field smoothed over a
`galactic' scale, $k_{G}=1h{\rm Mpc}^{-1}$, (using a sharp $k$-space
filter) is $B_{0}$. This gives, (for $n>-3$ {and for $k<k_c$})
\EQ
\Delta _{b}^{2}(k)= \frac{k^3M(k)}{2\pi^2}
=\frac{B_0^2}{2}(n+3)\left(\frac{k}{k_{G}}\right)^{3+n}.
\EN

The $3$-point correlation function of $\hat\Omega_B(\kk)$ involves a $6$-point correlation function of the fields.
A long but tedious calculation gives
$\zeta_{_{123}} = 
\delta(\kk_1 + \kk_2 + \kk_3) ~ \psi_{_{123}}$, where
\EQ
\psi_{_{123}} = \frac{1}{(4\pi\rho_0)^3}
\int d^3s M(\vert \kk_1 + \sss \vert) M(s) M(\vert \sss -\kk_3 \vert) F .
\label{psi}
\EN
Here $F = \alpha^2 + \beta^2 +\gamma^2 - \alpha\beta\gamma$ with
$\alpha = (\widehat{\sss}\cdot\widehat{\sss +\kk_1})$, 
$\beta = (\widehat{\sss}\cdot\widehat{\sss -\kk_3})$
and $\gamma = (\widehat{\kk_1 + \sss}\cdot\widehat{\sss -\kk_3})$,
where the hat on a vector denotes its unit vector.
This result has also been obtained in Ref. \cite{BC05}.

We calculate the bispectrum in two limits: 
(i) the `equilateral' case
for which the three $l_i$'s are equal, and,
(ii) the `local isosceles' case for which
$l_2 = l_3 \gg l_1$. In the former case, the presence 
of $j_{_{l_i}}(k_iD^*)$ in \Eq{bispec}, predominantly picks out 
configurations in wave-number space, in the $\Omega_B$ bispectrum,
with all the $k$'s are almost equal. In the latter case, it picks 
out configurations in 
$k$ values with $k_2 \sim k_3 \gg k_1$.

The mode coupling integral can be usefully approximated in these
two limits following methods discussed in earlier works
\cite{trsks01,SSB03}. In case (i), when $k_1=k_2=k_3$, we split 
the $s$-integral into the sub-ranges $0<s<k_1$ and $s>k_1$. 
In each of these sub-ranges we approximate the mode-coupling 
integrand in Eq.~\ref{psi}
by assuming that $s \ll k_1$ and $s \gg k_1$, respectively.
Similarly, in the second case, when $k_2= k_3 \gg k_1$, the
$s$-integral is now split into sub-ranges, $0 < s < k_1$,
$k_1 < s < k_3$ and $s > k_3$, and again in each of these sub-ranges,
we approximate the mode coupling integrand in Eq.~\ref{psi}
by assuming that
$s \ll k_1$, $k_1 \ll s \ll k_3$ and $s \gg k_3$, respectively.
We also restrict ourselves to spectral indices
$ -3 < n < -3/2$. In fact blue spectra are strongly constrained
by a number of observations, particularly the gravitational wave
limits of Ref. \cite{cap}. For numerical
estimates we will focus on nearly scale invariant spectra, that is
$n \to -3$. For case (i) we then get 

\EQ
\psi_{123} = \left(\frac{4}{3}\right)^4 \frac{\pi^7}{k_G^6}
\frac{(n+3)^2(7-n)}{2(\vert n +1\vert)}
\left(\frac{k_1}{k_G}\right)^{2n+3} \left(\frac{k_3}{k_G}\right)^n V_A^6,
\label{psi_approx2}
\EN

while for case (ii) we have

\EQ
\psi_{123} = \left(\frac{16}{3}\right)^3 \frac{\pi^7}{k_G^6}
\frac{(n+3)^2}{\vert 2n +3\vert}
\left(\frac{k_1}{k_G}\right)^{2n+3} \left(\frac{k_3}{k_G}\right)^n V_A^6. 
\label{psi_approx}
\EN
Here we have defined $V_A$, the Alfv\'en velocity in the radiation era as
\EQ
V_{A}={\frac{B_{0}}{(16\pi \rho _{0}/3)^{1/2}}}\approx 3.8\times
10^{-4}B_{-9},  
\label{alfvel}
\EN
with $B_{-9}\ \equiv (B_{0}/10^{-9}{\rm Gauss})$.

We now express the delta function in its integral form
$\delta(\kk) = (1/(2\pi)^3)\int d^3x \exp(i \kk\cdot\xx)$,
use the spherical wave expansion of the exponential terms,
substitute it into \Eq{bispec}, and integrate over the angular parts of
$(\kk_1,\kk_2,\kk_3,\xx)$. 
This algebra is very similar to what is done for calculating
the primordial bispectrum \cite{FS07}.
After this algebra we can write the
bispectrum $B^{m_{_1}m_{_2}m_{_3}}_{l_{_1}l_{_2}l_{_3}}$, 
in terms of a reduced bispectrum $b_{l_{_1}l_{_2}l_{_3}}$ as
\EQ
B^{m_{_1}m_{_2}m_{_3}}_{l_{_1}l_{_2}l_{_3}} =
{\cal G}_{m_{_1}m_{_2}m_{_3}}^{l_{_1}l_{_2}l_{_3}}
~ b_{l_{_1}l_{_2}l_{_3}}
\label{gaunt1}
\EN
where
\EQA
b_{l_{_1}l_{_2}l_{_3}} &=& 
\left(\frac{{\cal R}}{\pi^2}\right)^3 \int x^2 dx 
\nonumber \\
&\times&\prod_{i=1}^3
\int k_i^2dk_i ~ j_{_{l_i}}(k_{_i}x) ~ j_{_{l_i}}(k_{_i}D^*)
~ \psi_{_{123}}
\label{redbispec}
\ENA
and we have introduced the Gaunt integral
\EQ
{\cal G}_{m_{_1}m_{_2}m_{_3}}^{l_{_1}l_{_2}l_{_3}}
= \int d\Omega ~ Y_{l_1m_1}Y_{l_2m_2}Y_{l_3m_3} .
\label{gaunt}
\EN

For case (i), we substitute \Eq{psi_approx2} 
into \Eq{redbispec} for the reduced bispectrum.
Similarly for case (ii) we substitute \Eq{psi_approx}
into \Eq{redbispec}.
The integrals over $k_2$ can  be immediately done using
$\int k_2^2 dk_2 j_{l_2}(k_2x) j_{l_2}(k_2D^*) = (\pi/2x^2) \delta(x - D^*)$,
and the delta function makes the $x$-integral trivial. We are
then left with integrals over $k_1$ and $k_3$ given by
\EQA
b_{l_{_1}l_{_2}l_{_3}} &=& \frac{\pi}{2}
\left(\frac{{\cal R}}{\pi^2}\right)^3 
V_A^6 
\left[\int \frac{dk_3}{k_3} j_{_{l_3}}^2(k_3D^*) 
\left(\frac{k_3}{k_G}\right)^{n+3}
\right]
\nonumber \\
&\times& 
\left[\int \frac{dk_1}{k_1}j_{_{l_1}}^2(k_1D^*) 
\left(\frac{k_1}{k_G}\right)^{2(n+3)}
\right]
~ C(n). 
\label{bispec_kappa}
\ENA
Here for case (i) 
\EQ
C(n) = \left(\frac{4}{3}\right)^4 \frac{\pi^7}{2}
\frac{(n+3)^2(7-n)}{(\vert n +1\vert)}
\label{cn1}
\EN
where as for case (ii)
\EQ
C(n) = \left(\frac{16}{3}\right)^3 \pi^7
\frac{(n+3)^2}{\vert 2n +3\vert}.
\label{cn2}
\EN
Note that for any $n$ we can evaluate the integral in
\Eq{bispec_kappa} analytically in terms of Gamma functions. 
For power law spectra, the form of the integrals is the same
as the usual Sachs-Wolfe term. 
We focus on the nearly scale invariant case, $n \approx -3$.
Such magnetic spectra are expected to arise in inflationary models
for primordial magnetic field generation that we assume here
\cite{TW88}.
Further, let us first consider the purely
Sachs-Wolfe contribution as in \cite{giovannini}, which
gives ${\cal R} = R_\gamma/20 \sim 0.03$.
For the equilateral case we then have
\EQ
l_1(l_1+1)l_3(l_3+1) b_{l_{_1}l_{_2}l_{_3}}
\approx 2.3 \times 10^{-23} \left(\frac{n+3}{0.2}\right)^2
\left(\frac{B_{-9}}{3}\right)^6
\label{bispec_result1}
\EN
while for the local-isosceles case we get
\EQ
l_1(l_1+1)l_3(l_3+1) b_{l_{_1}l_{_2}l_{_3}}
\approx 1.5 \times 10^{-22} \left(\frac{n+3}{0.2}\right)^2 
\left(\frac{B_{-9}}{3}\right)^6.
\label{bispec_result2}
\EN
{
Including an ISW contribution can lead to larger signals.
We remark that for $n \sim -3$, the numerical values 
of $l_1(l_1+1)l_3(l_3+1) b_{l_{_1}l_{_2}l_{_3}}$
is higher for the local isosceles case by a factor $\sim 6.4$,
compared to the the equilateral case. 
The relative strengths of the bispectrum for the equilateral and local case has its origin in the ratio of $\psi_{123}$ for these two cases.} It can indeed be seen from equations \ref{psi_approx2} and \ref{psi_approx} that 
${\psi_{123(local)}}={\psi_{123(equil)}}(96 \mid n+1 \mid)/(\mid 2n+3\mid (7-n))$
which for $n \sim -3$ comes out to be $\sim 6.4 {\psi_{123(equil)}}$.

These values for the reduced bispectrum, 
should be compared with a value
$l_1(l_1+1)l_3(l_3+1) b_{l_{_1}l_{_2}l_{_3}} \sim 4 \times 10^{-18} f_{NL}$
at large angular scales, which arises due 
to nonlinear terms in the gravitational potential,
characterized by $f_{NL}$ (cf. \cite{riotto}).
Thus the magnetically induced non-Gaussian signal,
due to the purely Sachs-Wolfe effect  
(with ${\cal R}$ as in \cite{giovannini}),
is a factor of about  a {few times $10^4$}
smaller than the standard signal predicted in 
inflationary models with $f_{NL} \sim1$.
Conversely if observations constrain 
$l_1(l_1+1)l_3(l_3+1) b_{l_{_1}l_{_2}l_{_3}} < 4 \times 10^{-16}$,
assuming $f_{NL} < 100$ (for example say from WMAP 
experiment \cite{komatsu,wmap}),
then we have a limit of $B_0 < 35$ nano Gauss 
on the strength of any primordial field
with a nearly scale invariant spectrum.
Stronger limits 
would be obtained if we were to use 
potentially larger contributions from the ISW effect.

We note that the $l_i$ dependence for the standard  primordial
contribution to the bispectrum on large angular scales
(for scale invariant potential perturbations) is the same as
that we get here, 
in both cases (i) and (ii),
for magnetically induced
Sachs-Wolfe effect (for scale-invariant magnetic spectra)
(cf. \cite{FS07,riotto}).
That is in all these cases $l_1(l_1+1)l_3(l_3+1) b_{l_{_1}l_{_2}l_{_3}}$
is independent of $l$'s (see Eq. (17) and (18) and Ref.~\cite{riotto}).
Thus the bounds on $f_{NL}$ got from searching the WMAP data
for the standard `local' non-Gaussianity are indeed useful 
to set constraints on $B_0$. The WMAP limits use the much larger 
range of $l$ values than the range for which the SW contribution 
is important. However the limit on $B_0$ depends only very weakly 
($ B_0 \propto f_{NL}^{1/6}$) on the exact observational limit on $f_{NL}$.
Thus our limits on $B_0$ are expected to be reasonably robust.

A systematic, statistical study of limits 
on $B_0$ from the CMB power spectrum alone, taking account of all the available
data and comparing with the sum of scalar, vector and tensor modes of the
magnetically induced signals, although challenging, would be of great interest.
Present works deal with only a subset of the data or the signals:
In typical models a field $B_0 \sim 3$ nG for scale invariant spectra, are
not excluded by any current CMB data (cf. \cite{Bcmb,trsks01,SSB03,finelli} 
and the reviews \cite{reviews}),
while $B_0 \sim 10$ nG would seem to be excluded.
In this paper we have attempted to probe stochastic primordial
magnetic fields, arising from
SW effect. The present constraint on $B_0$ derived above from the non-Gaussianity limits
arises from the magnetically induced SW contribution to the bispectrum. Although the SW contribution above is weaker than that expected from inflationary contribution, it is of interest as this is a  novel probe of stochastic primordial
magnetic fields. This is of value for setting the stage for calculating the full contribution by including vector as well tensor modes. Vector modes are expected to contribute on small angular scales (large $l$). Tensor modes could contribute over the same range of $l$-values considered here. Adding these contributions is expected to lead to stronger limits on $B_0$ 
from limits on CMB non-Gaussianity.
It is worth noting that if only the magnetically induced SW contribution is taken into account even the limit on $B_0$ from the CMB power spectrum would be much weaker.
We have focused
on SW contribution of scalar modes here so as to calculate, as a first cut,
the simplest contribution
to non-Gaussian signals induced by stochastic primordial magnetic fields.

There is a distinct advantage of using bispectrum (as a probe for stochastic primordial magnetic fields)
over the power spectrum. The magnetically induced
signal being fundamentally non Gaussian, 
could be more easily distinguishable
in the bispectrum. This is due to the fact that bispectrum arising due to magnetic contribution
can in principle dominate that arising from,
models of inflation with a small enough $f_{NL}$.
On the other hand, the problem with the power spectrum arising due to magnetic contribution is that it is generally subdominant to those arising from
inflation generated curvature perturbations, for nano Gauss fields
and scale invariant spectra. These will hence require careful analysis to isolate the
magnetic contribution. 
The probe using bispectrum developed in this paper is free of such limitations.

In conclusion we have studied a new probe of primordial
magnetic fields, by computing the CMB bispectrum
they induce on large angular scales.
The magnetically induced signals on the CMB anisotropies
have a necessarily a non-Gaussian character, as
the temperature fluctuations they induce depend quadratically
on the field strength. 
The CMB bispectrum is an important and useful probe
of this non-Gaussian nature.
Earlier work has emphasized the
role of the CMB angular power spectrum in
constraining and detecting primordial fields.
As we have argued here, 
the bispectrum will provide a new and independent handle
for constraining such fields. 
Further numerical work,
which also considers 
all the scalar, 
vector and tensor modes, will
help to strengthen the constraints derived here. 

{\em Note added:} After we had submitted this paper a preprint \cite{caprini} appeared which confirms our main conclusions.

{\em Acknowledgement:} The authors thank the anonymous referees for their many useful suggestions that have helped in improving this paper.


\begin{thebibliography}{99}
\bibitem{wmap} 
The WMAP 5-year data gives limits on primordial non-Gaussianity 
$-9<f_{NL}^{local}<111$ and $-151<f_{NL}^{equil}<253$ at 
the $95\%$ confidence level
for the local and equilateral models, respectively. 
Of course these limits use the much 
larger range of $l$ values than the range for which the SW contribution is important.
\end{thebibliography}
\end{document}